\documentclass[10pt,a4paper,onecolumn]{article}
\usepackage{marginnote}
\usepackage{graphicx}
\usepackage{xcolor}
\usepackage{authblk,etoolbox}
\usepackage{titlesec}
\usepackage{calc}
\usepackage{tikz}
\usepackage{hyperref}
\hypersetup{colorlinks,breaklinks=true,
            urlcolor=[rgb]{0.0, 0.5, 1.0},
            linkcolor=[rgb]{0.0, 0.5, 1.0}}
\usepackage{caption}
\usepackage{tcolorbox}
\usepackage{amssymb,amsmath}
\usepackage{ifxetex,ifluatex}
\usepackage{seqsplit}
\usepackage{xstring}

\usepackage{float}
\let\origfigure\figure
\let\endorigfigure\endfigure
\renewenvironment{figure}[1][2] {
    \expandafter\origfigure\expandafter[H]
} {
    \endorigfigure
}

\usepackage{fixltx2e} 
\usepackage[
  backend=biber,
]{biblatex}
\bibliography{paper.bib}

\let\textttOrig=\texttt
\def\texttt#1{\expandafter\textttOrig{\seqsplit{#1}}}
\renewcommand{\seqinsert}{\ifmmode
  \allowbreak
  \else\penalty6000\hspace{0pt plus 0.02em}\fi}

\makeatletter
\let\href@Orig=\href
\def\href@Urllike#1#2{\href@Orig{#1}{\begingroup
    \def\Url@String{#2}\Url@FormatString
    \endgroup}}
\def\href@Notdoi#1#2{\def\tempa{#1}\def\tempb{#2}%
  \ifx\tempa\tempb\relax\href@Urllike{#1}{#2}\else
  \href@Orig{#1}{#2}\fi}
\def\href#1#2{%
  \IfBeginWith{#1}{https://doi.org}%
  {\href@Urllike{#1}{#2}}{\href@Notdoi{#1}{#2}}}
\makeatother

\newlength{\cslhangindent}
\newlength{\csllabelwidth}
\newenvironment{CSLReferences}[3] 
 {
  \setlength{\parindent}{0pt}
  \ifodd #1 \everypar{\setlength{\hangindent}{\cslhangindent}}\ignorespaces\fi
  \ifnum #2 > 0
  \setlength{\parskip}{#2\baselineskip}
  \fi
 }%
 {}
\usepackage{calc}

\usepackage[top=3.5cm, bottom=3cm, right=1.5cm, left=1.0cm,
            headheight=2.2cm, reversemp, includemp, marginparwidth=4.5cm]{geometry}

\titleformat{\section}
  {\normalfont\sffamily\Large\bfseries}
  {}{0pt}{}
\titleformat{\subsection}
  {\normalfont\sffamily\large\bfseries}
  {}{0pt}{}
\titleformat{\subsubsection}
  {\normalfont\sffamily\bfseries}
  {}{0pt}{}
\titleformat*{\paragraph}
  {\sffamily\normalsize}

\usepackage{fancyhdr}
\makeatletter
\let\ps@plain\ps@fancy
\fancyheadoffset[L]{4.5cm}
\fancyfootoffset[L]{4.5cm}

\definecolor{linky}{rgb}{0.0, 0.5, 1.0}

\newtcolorbox{repobox}
   {colback=red, colframe=red!75!black,
     boxrule=0.5pt, arc=2pt, left=6pt, right=6pt, top=3pt, bottom=3pt}

\newcommand{\ExternalLink}{%
   \tikz[x=1.2ex, y=1.2ex, baseline=-0.05ex]{%
       \begin{scope}[x=1ex, y=1ex]
           \clip (-0.1,-0.1)
               --++ (-0, 1.2)
               --++ (0.6, 0)
               --++ (0, -0.6)
               --++ (0.6, 0)
               --++ (0, -1);
           \path[draw,
               line width = 0.5,
               rounded corners=0.5]
               (0,0) rectangle (1,1);
       \end{scope}
       \path[draw, line width = 0.5] (0.5, 0.5)
           -- (1, 1);
       \path[draw, line width = 0.5] (0.6, 1)
           -- (1, 1) -- (1, 0.6);
       }
   }

\patchcmd{\@maketitle}{center}{flushleft}{}{}
\patchcmd{\@maketitle}{center}{flushleft}{}{}
\patchcmd{\@maketitle}{\LARGE}{\LARGE\sffamily}{}{}
\def\maketitle{{%
  
  \AB@maketitle}}
\makeatletter
\renewcommand\AB@affilsepx{ \protect\Affilfont}
\renewcommand\AB@affilnote[1]{{\bfseries #1}\hspace{3pt}}
\renewcommand{\affil}[2][]%
   {\newaffiltrue\let\AB@blk@and\AB@pand
      \if\relax#1\relax\def\AB@note{\AB@thenote}\else\def\AB@note{#1}%
        \setcounter{Maxaffil}{0}\fi
        \begingroup
        \let\href=\href@Orig
        \let\texttt=\textttOrig
        \let\protect\@unexpandable@protect
        \def\thanks{\protect\thanks}\def\footnote{\protect\footnote}%
        \@temptokena=\expandafter{\AB@authors}%
        {\def\\{\protect\\\protect\Affilfont}\xdef\AB@temp{#2}}%
         \xdef\AB@authors{\the\@temptokena\AB@las\AB@au@str
         \protect\\[\affilsep]\protect\Affilfont\AB@temp}%
         \gdef\AB@las{}\gdef\AB@au@str{}%
        {\def\\{, \ignorespaces}\xdef\AB@temp{#2}}%
        \@temptokena=\expandafter{\AB@affillist}%
        \xdef\AB@affillist{\the\@temptokena \AB@affilsep
          \AB@affilnote{\AB@note}\protect\Affilfont\AB@temp}%
      \endgroup
       \let\AB@affilsep\AB@affilsepx
}
\makeatother

\renewcommand\Affilfont{\sffamily\small\mdseries}
\ifnum 0\ifxetex 1\fi\ifluatex 1\fi=0 
  \usepackage[T1]{fontenc}
  \usepackage[utf8]{inputenc}

\else 
  \ifxetex
    \usepackage{mathspec}
    \usepackage{fontspec}

  \else
    \usepackage{fontspec}
  \fi
  \defaultfontfeatures{Ligatures=TeX,Scale=MatchLowercase}

\fi
\IfFileExists{upquote.sty}{\usepackage{upquote}}{}
\IfFileExists{microtype.sty}{%
\usepackage{microtype}
\UseMicrotypeSet[protrusion]{basicmath} 
}{}

\usepackage{hyperref}
\hypersetup{unicode=true,
            pdftitle={Interface to high-performance periodic coupled-cluster theory calculations with atom-centered, localized basis functions},
            pdfborder={0 0 0},
            breaklinks=true}
\let\addcontentslineOrig=\addcontentsline
\def\addcontentsline#1#2#3{\bgroup
  \let\texttt=\textttOrig\addcontentslineOrig{#1}{#2}{#3}\egroup}
\let\markbothOrig\markboth
\def\markboth#1#2{\bgroup
  \let\texttt=\textttOrig\markbothOrig{#1}{#2}\egroup}
\let\markrightOrig\markright
\def\markright#1{\bgroup
  \let\texttt=\textttOrig\markrightOrig{#1}\egroup}

\IfFileExists{parskip.sty}{%
\usepackage{parskip}
}{
\setlength{\parindent}{0pt}
\setlength{\parskip}{6pt plus 2pt minus 1pt}
}
\ifx\paragraph\undefined\else
\let\oldparagraph\paragraph
\renewcommand{\paragraph}[1]{\oldparagraph{#1}\mbox{}}
\fi
\ifx\subparagraph\undefined\else
\let\oldsubparagraph\subparagraph
\renewcommand{\subparagraph}[1]{\oldsubparagraph{#1}\mbox{}}
\fi

\title{Interface to high-performance periodic coupled-cluster theory
calculations with atom-centered, localized basis functions}

        \author[1]{Evgeny Moerman}
          \author[2]{Felix Hummel}
          \author[2]{Andreas Grüneis}
          \author[2]{Andreas Irmler}
          \author[1]{Matthias Scheffler}
    
      \affil[1]{The NOMAD Laboratory at the Fritz Haber Institute of the
Max Plank Society, Berlin, Germany}
      \affil[2]{Institute for Theoretical Physics, TU Wien, Vienna,
Austria}
  \date{\vspace{-7ex}}

\begin{document}
\maketitle

\marginpar{

  \begin{flushleft}
  \sffamily\small

  {\bfseries DOI:} \href{https://doi.org/DOI unavailable}{\color{linky}{DOI unavailable}}

  \vspace{2mm}

  {\bfseries Software}
  \begin{itemize}
    \setlength\itemsep{0em}
    \item \href{https://github.com/openjournals/joss-reviews/issues/4040}{\color{linky}{Review}} \ExternalLink
    \item \href{https://gitlab.com/moerman1/fhi-cc4s}{\color{linky}{Repository}} \ExternalLink
    \item \href{DOI unavailable}{\color{linky}{Archive}} \ExternalLink
  \end{itemize}

  \vspace{2mm}

  \par\noindent\hrulefill\par

  \vspace{2mm}

  {\bfseries Editor:} \href{https://example.com}{Pending
Editor} \ExternalLink \\
  \vspace{1mm}
    {\bfseries Reviewers:}
  \begin{itemize}
  \setlength\itemsep{0em}
    \item \href{https://github.com/Pending Reviewers}{@Pending
Reviewers}
    \end{itemize}
    \vspace{2mm}

  {\bfseries Submitted:} N/A\\
  {\bfseries Published:} N/A

  \vspace{2mm}
  {\bfseries License}\\
  Authors of papers retain copyright and release the work under a Creative Commons Attribution 4.0 International License (\href{http://creativecommons.org/licenses/by/4.0/}{\color{linky}{CC BY 4.0}}).

  \end{flushleft}
}

\hypertarget{summary}{%
\section{Summary}\label{summary}}

A major part of computational materials science and computational
chemistry concerns calculations of total energy differences and
electronic excitations of poly-atomic systems. Currently, the most
prevalent method for such computations is density-functional theory
(Hohenberg \& Kohn, 1964) (DFT) based on the Kohn-Sham formalism (Kohn
\& Sham, 1965) (KS) together with one of the numerous
exchange-correlation (xc) approximations (Civalleri et al., 2012; Sousa
et al., 2007). The trade-off between comparably low computational cost
and often reliably accurate results render it the dominant method in the
field. Often, however, the accuracy of the xc approximations is not
sufficent and uncertain, in particular when electronic correlations play
a decisive role (Savin \& Johnson, 2014; Zhang et al., 2016; Zhang \&
Grüneis, 2019).\\
Coupled-cluster (CC) methods (Čížek, 1966), while substantially more
computationally expensive have proven to be significantly more accurate
and reliable, at least for molecules and non-metallic solids. On these
grounds, they allow for a systematic accuracy benchmark of other
methods. Indeed, in molecular quantum chemistry the CC approach is
considered the gold standard for a theoretical description of binding
energies and electronic properties (Chan, 2019). It is typically used
(at least) for benchmark studies. While the original CC methodology
allows to calculate the groundstate total energy of a system, it is also
possible to compute properties of excited states using the
equation-of-motion formalism of CC (EOM-CC) with comparable accuracy and
reliability (Stanton \& Bartlett, 1993). The conventional CC approach is
limited, however, to systems with about 50 electrons (Feller \& Dixon,
2001; Gyevi-Nagy et al., 2019). By employing approximations, which
exploit the locality of electronic correlation, materials with a couple
of hundreds of electrons have been calculated via local natural orbital
CC (LNO-CC) (Nagy \& Kállay, 2019) and explicitly correlated pair
natural orbital CC (PNO-CC-F12) (Ma \& Werner, 2021). The extension of
CC to periodic systems (Hirata et al., 2004) has been explored to a
great degree by the research groups of
\href{http://cqc.itp.tuwien.ac.at/code.html}{Andreas Grüneis} and
\href{https://pyscf.org/}{Garnet Chan}.

In the work of Andreas Grüneis et al.~the periodic formulation of CC
was, for example, applied to study the adsorption behavior and surface
chemistry of 2-dimensional materials including graphene (Al-Hamdani et
al., 2017), boron-nitride (Brandenburg et al., 2019) and surfaces
(Tsatsoulis et al., 2018). These studies showed that CC yields
consistently accurate adsorption energies and reaction energetics. The
work of the groups of Garnet Chan and Timothy Berkelbach addressed the
electronic properties of 2- and 3-dimensional materials. By applying the
periodic equation-of-motion (EOM) CC formalism, band paths, optical
spectra and band gaps of various materials (diamond, silicon, nickel
oxide and others)\\
have been investigated (Gao et al., 2020; McClain et al., 2017; Wang \&
Berkelbach, 2020).

It is important to note that all the periodic CC calculations published
so far utilized pseudopotentials (Kresse \& Joubert, 1999) to largely
disregard the effect of core electrons. One notable distinction between
the works of the two groups, is the type of basis set used. While the
\href{http://cqc.itp.tuwien.ac.at/code.html}{CC4S} code of Andreas
Grüneis et al.~uses a plane-wave basis, the group of Garnet Chan perform
their calculations using gaussian basis sets (McClain et al., 2017; Sun
et al., 2018). Localized atom-centred basis sets like gaussian orbitals
or numerical atomic orbitals (NAO) (Blum et al., 2009) can potentially
decrease the computational cost. This is mostly due to the locality of
the basis functions and their improved description of the atomic core
region, which decreases the number of basis functions necessary for
accurate computations of the system.

This paper describes a generalizable interface, called \texttt{CC-aims},
to the \texttt{Coupled-Cluster\ for\ solids} code
(\href{http://cqc.itp.tuwien.ac.at/code.html}{CC4S}) developed by
Andreas Grüneis et al. The interface is formulated in a general way and
demonstrated here for the all-electron FHI-aims code (Blum et al.,
2009). A generalization to other codes with atom-centered basis
functions is straight forward. This interface expands the power of
electronic-structure theory codes to a variety of correlated methods.
These include Møller-Plesset perturbation theory to second order (MP2),
coupled-cluster theory including single and double excitations (CCSD)
and CCSD including the perturbative treatment of triple excitations
(CCSD(T)). Implementations of EOM-CC for neutral (EE-EOM-CC) and charged
(IP-EOM-CC/EA-EOM-CC) excitations are currently under development.
\texttt{CC-aims} can be used directly by any software package which uses
a localized basis set and employs a resolution-of-identity scheme (Ren
et al., 2012) (RI). On the one hand, this includes local RI schemes,
which expand products of atomic orbital basis function (AOs) pairs only
using auxiliary basis functions which are localized on either of the
atoms of the AOs. Primary examples for this family of localized schemes
is the RI-LVL approach employed by FHI-aims (Ihrig et al., 2015), ADF
(Förster \& Visscher, 2020) and ABACUS (Lin et al., 2020) and the the
pair-atomic RI (PARI) (Merlot et al., 2013). On the other hand, more
conventional non-local schemes, which are predominantly used in
molecular calculations, like RI-V (Whitten, 1973) and RI-SVS (Feyereisen
et al., 1993) are recognized by CC-aims as well.

\hypertarget{statement-of-need}{%
\section{Statement of need}\label{statement-of-need}}

The only input quantity of CC4S, which is not typically calculated in
quantum chemistry or electronic structure codes, but which is computed
by \texttt{CC-aims}, is the Coulomb vertex (Hummel et al., 2017). The
Coulomb vertex, a rank-3 tensor, constitutes a memory-saving,
approximation of the rank-4 tensor of Coulomb integrals. The storage of
Coulomb integrals is the major memory bottleneck in CC calculations, so
that the utilization of the Coulomb vertex expands the scope of system
sizes which can be calculated. While in the case of localized orbitals
the herein presented CC-aims interface can be used, in the case of
plane-wave basis sets a different approach to calculate the Coulomb
vertex has to be taken, which is described in (Hummel et al., 2017). For
Quantum chemistry programs employing a localized basis additionally the
use of an RI scheme is needed or needs to be implemented.
\texttt{CC-aims} allows software packages which either lack certain
Quantum chemistry algorithms completely or which only offer
insufficiently optimized implementations easy access to these methods.
Interfaces like \texttt{CC-aims} will substantially accelerate the
research done in areas, where DFT is too inaccurate or too unreliable,
by allowing many electronic structure codes to participate in these
investigations without the time-consuming effort of implementing
correlated wave-function methods.

\hypertarget{acknowledgement}{%
\section{Acknowledgement}\label{acknowledgement}}

This work received funding from the European Union's Horizon 2020
Research and Innovation Programme (grant agreement No.~951786, the NOMAD
CoE), and the ERC Advanced Grant TEC1P (No.~740233)

\hypertarget{references}{%
\section*{References}\label{references}}
\addcontentsline{toc}{section}{References}

\hypertarget{refs}{}
\begin{CSLReferences}{1}{0}
\leavevmode\hypertarget{ref-Al:2017}{}%
Al-Hamdani, Y. S., Rossi, M., Alfe, D., Tsatsoulis, T., Ramberger, B.,
Brandenburg, J. G., Zen, A., Kresse, G., Grüneis, A., Tkatchenko, A., \&
others. (2017). Properties of the water to boron nitride interaction:
From zero to two dimensions with benchmark accuracy. \emph{The Journal
of Chemical Physics}, \emph{147}(4), 044710.
\url{https://doi.org/10.1063/1.4985878}

\leavevmode\hypertarget{ref-Blum:2009}{}%
Blum, V., Gehrke, R., Hanke, F., Havu, P., Havu, V., Ren, X., Reuter,
K., \& Scheffler, M. (2009). Ab initio molecular simulations with
numeric atom-centered orbitals. \emph{Computer Physics Communications},
\emph{180}(11), 2175--2196.
\url{https://doi.org/10.1016/j.cpc.2009.06.022}

\leavevmode\hypertarget{ref-Brandenburg:2019}{}%
Brandenburg, J. G., Zen, A., Fitzner, M., Ramberger, B., Kresse, G.,
Tsatsoulis, T., Grüneis, A., Michaelides, A., \& Alfè, D. (2019).
Physisorption of water on graphene: Subchemical accuracy from many-body
electronic structure methods. \emph{The Journal of Physical Chemistry
Letters}, \emph{10}(3), 358--368.
\url{https://doi.org/10.1021/acs.jpclett.8b03679}

\leavevmode\hypertarget{ref-Chan:2019}{}%
Chan, B. (2019). The CUAGAU set of coupled-cluster reference data for
small copper, silver, and gold compounds and assessment of DFT methods.
\emph{The Journal of Physical Chemistry A}, \emph{123}(27), 5781--5788.
\url{https://doi.org/10.1021/acs.jpca.9b03976}

\leavevmode\hypertarget{ref-Civalleri:2012}{}%
Civalleri, B., Presti, D., Dovesi, R., \& Savin, A. (2012). On choosing
the best density functional approximation. \emph{Chem. Modell},
\emph{9}, 168--185. \url{https://doi.org/10.1039/9781849734790-00168}

\leavevmode\hypertarget{ref-Cizek:1966}{}%
Čížek, J. (1966). On the correlation problem in atomic and molecular
systems. Calculation of wavefunction components in ursell-type expansion
using quantum-field theoretical methods. \emph{The Journal of Chemical
Physics}, \emph{45}(11), 4256--4266.

\leavevmode\hypertarget{ref-Feller:2001}{}%
Feller, D., \& Dixon, D. A. (2001). Extended benchmark studies of
coupled cluster theory through triple excitations. \emph{The Journal of
Chemical Physics}, \emph{115}(8), 3484--3496.
\url{https://doi.org/10.1063/1.1388045}

\leavevmode\hypertarget{ref-Feyereisen:1993}{}%
Feyereisen, M., Fitzgerald, G., \& Komornicki, A. (1993). Use of
approximate integrals in ab initio theory. An application in MP2 energy
calculations. \emph{Chemical Physics Letters}, \emph{208}(5-6),
359--363. \url{https://doi.org/10.1016/0009-2614(93)87156-w}

\leavevmode\hypertarget{ref-Forster:2020}{}%
Förster, A., \& Visscher, L. (2020). Low-order scaling g 0 w 0 by pair
atomic density fitting. \emph{Journal of Chemical Theory and
Computation}, \emph{16}(12), 7381--7399.

\leavevmode\hypertarget{ref-Gao:2020}{}%
Gao, Y., Sun, Q., Jason, M. Y., Motta, M., McClain, J., White, A. F.,
Minnich, A. J., \& Chan, G. K.-L. (2020). Electronic structure of bulk
manganese oxide and nickel oxide from coupled cluster theory.
\emph{Physical Review B}, \emph{101}(16), 165138.
\url{https://doi.org/10.1103/physrevb.101.165138}

\leavevmode\hypertarget{ref-Gyevi:2019}{}%
Gyevi-Nagy, L., Kállay, M., \& Nagy, P. R. (2019). Integral-direct and
parallel implementation of the CCSD (t) method: Algorithmic developments
and large-scale applications. \emph{Journal of Chemical Theory and
Computation}, \emph{16}(1), 366--384.
\url{https://doi.org/10.1021/acs.jctc.9b00957}

\leavevmode\hypertarget{ref-Hirata:2004}{}%
Hirata, S., Podeszwa, R., Tobita, M., \& Bartlett, R. J. (2004).
Coupled-cluster singles and doubles for extended systems. \emph{The
Journal of Chemical Physics}, \emph{120}(6), 2581--2592.

\leavevmode\hypertarget{ref-Hohenberg:1964}{}%
Hohenberg, P., \& Kohn, W. (1964). Inhomogeneous electron gas.
\emph{Physical Review}, \emph{136}(3B), B864.

\leavevmode\hypertarget{ref-Hummel:2017}{}%
Hummel, F., Tsatsoulis, T., \& Grüneis, A. (2017). Low rank
factorization of the coulomb integrals for periodic coupled cluster
theory. \emph{The Journal of Chemical Physics}, \emph{146}(12), 124105.
\url{https://doi.org/10.1063/1.4977994}

\leavevmode\hypertarget{ref-Ihrig:2015}{}%
Ihrig, A. C., Wieferink, J., Zhang, I. Y., Ropo, M., Ren, X., Rinke, P.,
Scheffler, M., \& Blum, V. (2015). Accurate localized resolution of
identity approach for linear-scaling hybrid density functionals and for
many-body perturbation theory. \emph{New Journal of Physics},
\emph{17}(9), 093020.
\url{https://doi.org/10.1088/1367-2630/17/9/093020}

\leavevmode\hypertarget{ref-Kohn:1965}{}%
Kohn, W., \& Sham, L. J. (1965). Self-consistent equations including
exchange and correlation effects. \emph{Physical Review},
\emph{140}(4A), A1133.

\leavevmode\hypertarget{ref-Kresse:1999}{}%
Kresse, G., \& Joubert, D. (1999). From ultrasoft pseudopotentials to
the projector augmented-wave method. \emph{Physical Review b},
\emph{59}(3), 1758. \url{https://doi.org/10.1103/physrevb.59.1758}

\leavevmode\hypertarget{ref-Lin:2020}{}%
Lin, P., Ren, X., \& He, L. (2020). Accuracy of localized resolution of
the identity in periodic hybrid functional calculations with numerical
atomic orbitals. \emph{The Journal of Physical Chemistry Letters},
\emph{11}(8), 3082--3088.
\url{https://doi.org/10.1021/acs.jpclett.0c00481}

\leavevmode\hypertarget{ref-Ma:2021}{}%
Ma, Q., \& Werner, H.-J. (2021). Scalable electron correlation methods.
8. Explicitly correlated open-shell coupled-cluster with pair natural
orbitals PNO-RCCSD (t)-F12 and PNO-UCCSD (t)-F12. \emph{Journal of
Chemical Theory and Computation}, \emph{17}(2), 902--926.
\url{https://doi.org/10.1021/acs.jctc.0c01129}

\leavevmode\hypertarget{ref-Mcclain:2017}{}%
McClain, J., Sun, Q., Chan, G. K.-L., \& Berkelbach, T. C. (2017).
Gaussian-based coupled-cluster theory for the ground-state and band
structure of solids. \emph{Journal of Chemical Theory and Computation},
\emph{13}(3), 1209--1218. \url{https://doi.org/10.1021/acs.jctc.7b00049}

\leavevmode\hypertarget{ref-Merlot:2013}{}%
Merlot, P., Kjærgaard, T., Helgaker, T., Lindh, R., Aquilante, F.,
Reine, S., \& Pedersen, T. B. (2013). Attractive electron--electron
interactions within robust local fitting approximations. \emph{Journal
of Computational Chemistry}, \emph{34}(17), 1486--1496.
\url{https://doi.org/10.1002/jcc.23284}

\leavevmode\hypertarget{ref-Nagy:2019}{}%
Nagy, P. R., \& Kállay, M. (2019). Approaching the basis set limit of
CCSD (t) energies for large molecules with local natural orbital
coupled-cluster methods. \emph{Journal of Chemical Theory and
Computation}, \emph{15}(10), 5275--5298.
\url{https://doi.org/10.1021/acs.jctc.9b00511}

\leavevmode\hypertarget{ref-Ren:2012}{}%
Ren, X., Rinke, P., Blum, V., Wieferink, J., Tkatchenko, A., Sanfilippo,
A., Reuter, K., \& Scheffler, M. (2012). Resolution-of-identity approach
to hartree--fock, hybrid density functionals, RPA, MP2 and GW with
numeric atom-centered orbital basis functions. \emph{New Journal of
Physics}, \emph{14}(5), 053020.
\url{https://doi.org/10.1088/1367-2630/14/5/053020}

\leavevmode\hypertarget{ref-Savin:2014}{}%
Savin, A., \& Johnson, E. R. (2014). Judging density-functional
approximations: Some pitfalls of statistics. \emph{Density Functionals},
81--95. \url{https://doi.org/10.1007/128_2014_600}

\leavevmode\hypertarget{ref-Sousa:2007}{}%
Sousa, S. F., Fernandes, P. A., \& Ramos, M. J. (2007). General
performance of density functionals. \emph{The Journal of Physical
Chemistry A}, \emph{111}(42), 10439--10452.

\leavevmode\hypertarget{ref-Stanton:1993}{}%
Stanton, J. F., \& Bartlett, R. J. (1993). The equation of motion
coupled-cluster method. A systematic biorthogonal approach to molecular
excitation energies, transition probabilities, and excited state
properties. \emph{The Journal of Chemical Physics}, \emph{98}(9),
7029--7039. \url{https://doi.org/10.1063/1.464746}

\leavevmode\hypertarget{ref-Sun:2018}{}%
Sun, Q., Berkelbach, T. C., Blunt, N. S., Booth, G. H., Guo, S., Li, Z.,
Liu, J., McClain, J. D., Sayfutyarova, E. R., Sharma, S., \& others.
(2018). PySCF: The python-based simulations of chemistry framework.
\emph{Wiley Interdisciplinary Reviews: Computational Molecular Science},
\emph{8}(1), e1340.

\leavevmode\hypertarget{ref-Tsatsoulis:2018}{}%
Tsatsoulis, T., Sakong, S., Groß, A., \& Grüneis, A. (2018). Reaction
energetics of hydrogen on si (100) surface: A periodic many-electron
theory study. \emph{The Journal of Chemical Physics}, \emph{149}(24),
244105. \url{https://doi.org/10.1063/1.5055706}

\leavevmode\hypertarget{ref-Wang:2020}{}%
Wang, X., \& Berkelbach, T. C. (2020). Excitons in solids from periodic
equation-of-motion coupled-cluster theory. \emph{Journal of Chemical
Theory and Computation}, \emph{16}(5), 3095--3103.
\url{https://doi.org/10.1021/acs.jctc.0c00101}

\leavevmode\hypertarget{ref-Whitten:1973}{}%
Whitten, J. L. (1973). Coulombic potential energy integrals and
approximations. \emph{The Journal of Chemical Physics}, \emph{58}(10),
4496--4501. \url{https://doi.org/10.1063/1.1679012}

\leavevmode\hypertarget{ref-Zhang:2019}{}%
Zhang, I. Y., \& Grüneis, A. (2019). Coupled cluster theory in materials
science. \emph{Frontiers in Materials}, \emph{6}, 123.
\url{https://doi.org/10.3389/fmats.2019.00123}

\leavevmode\hypertarget{ref-Zhang:2016}{}%
Zhang, I. Y., Rinke, P., Perdew, J. P., \& Scheffler, M. (2016). Towards
efficient orbital-dependent density functionals for weak and strong
correlation. \emph{Physical Review Letters}, \emph{117}(13), 133002.
\url{https://doi.org/10.1103/physrevlett.117.133002}

\end{CSLReferences}

\end{document}